\documentclass[final]{aipproc}
  \layoutstyle{8x11double}

   \usepackage{graphicx}
   \usepackage{natbib}
   \usepackage{amssymb,amsmath}

   \newcommand{\be}{\begin{equation}}
   \newcommand{\ee}{\end{equation}}
   \newcommand{\ba}{\begin{eqnarray}}
   \newcommand{\ea}{\end{eqnarray}}
   
   \newcommand{\elsasser}{Els\"{a}sser }
   \newcommand{\alfven}{Alfv\'en }
   \newcommand{\dt}[1]{\frac{\partial #1}{\partial t}}

\begin{document}



\title{Coupling Photosphere and Corona: Linear and Turbulent Regimes}
               

\author{Verdini A.}{
  address={SIDC, Observatoire Royal de Belgique, Bruxelles, Belgium}
}
\author{Grappin R.}{
  address={LUTH, Observatoire de Paris, France}
}
\author{Velli M.}{
  address={Dipartimento di Astronomia e Scienza dello Spazio, Firenze,
        Italy}
  ,altaddress={Jet Propulsion Laboratory, California Institute of Technology,
        Pasadena, CA 91109, USA}
}


\begin{abstract}
In a recent work \citet{Grappin_al_2008} have shown that low- 
frequency movements can be transmitted from one footpoint to the other 
along a magnetic loop, thus mimicking a friction effect of the corona on the 
photosphere, and invalidating the line-tying approximation. 
We consider here successively the effect of high frequencies and turbulent 
damping on the process. We use a very simple atmospheric model which 
allows to study analytically the laminar case, and to study the turbulent case 
both using simple phenomenological arguments and a more sophisticated 
turbulence model 
\citep{Buchlin_Velli_2007}. 
We find that, except when turbulent damping is such that all turbulence is 
damped during loop traversal, coupling still occurs between distant 
footpoints, and moreover the coronal field induced by photospheric 
movements saturates at finite values.
\end{abstract}
\maketitle
In large simulations the line-tying approximation is usually 
adopted for the coronal boundary. According to this 
approximation, strong density gradients cause the 
reflection of any coronal disturbance. Hence, the 
photospheric velocity field can be prescribed since the reaction 
of the coronal dynamics is completely neglected. 
Applying a velocity field $u_0$ at the base of a loop of length $L$ 
and mean magnetic field $B_0$, keeping the other foot anchored 
in the photosphere leads to an 
infinite accumulation of magnetic energy (driven by the 
velocity shear between the loop footpoints).
To test whether the stratification really leads to line-tying, 
\citet{Grappin_al_2008} have studied the 
transmission of a signal from one footpoint to the other 
including a simple atmospheric model in a 1.5D MHD 
simulation. Footpoints are free to move, i.e. boundaries are 
transparent to waves, and a kick is given to left footpoint, injecting an
\alfven waves from the left boundary.
At the beginning, the magnetic energy grows steadily 
according to the line-tied assumption,
$b\approx B_0u_0t/L$ (transient acceleration of the left footpoint).
On longer times $\tau_{rel}=L/V_a^{ph}$
($V_a^{ph}$ is the \alfven speed at the photosphere), 
the leakage of \alfven waves through the transition region (T.R.)
accelerates (decelerates) the right (left) footpoint: 
the system relaxes to a state in which both footpoints have the same speed.
\section{Simplified model atmosphere}
\begin{figure}[t]
\centering
\includegraphics[width=1\columnwidth]{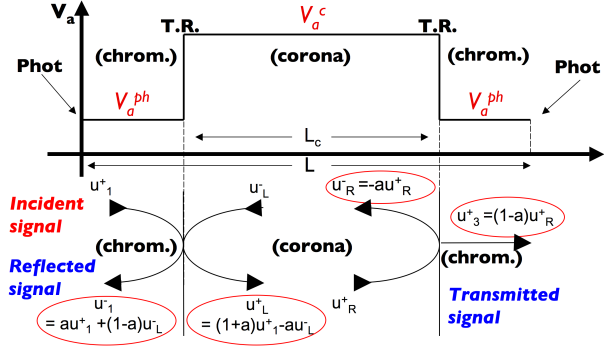}
\caption{Modeled loop atmosphere (top) and jump conditions (red circles) 
at the two T.R.s (bottom) for the rightward and leftward
propagating waves ($u^\pm$ respectively).}
\label{fig:1}
\end{figure}

We adopt a rough representation of the loop atmosphere modeling the transition
region as a discontinuity in the \alfven speed, 
separating two 
uniform \alfven speed chromosphere and corona 
(hereafter
subscript and/or superscript $c,~ph$ indicate quantities evaluated at the
corona and at the photosphere, not distinguished from the chromosphere in the
model).
This allows 
to easily consider the effect of finite frequency, wave reflection, and
wave damping while retaining the physical process responsible for the relaxation.\\
The small parameter $\epsilon=V_a^{ph}/V_a^c$ quantifies the jump at
the T.R. and hence reflection.
Continuity of magnetic and velocity field
fluctuation, $\delta b,~\delta u$, at the two T.R.s yields four
jump conditions for the \elsasser variables 
$u^\pm=\delta u \mp \delta b/\sqrt{4\pi\rho}$:
\ba
u^+_1 + u^-_1  =  u^+_L + u^-_L,
\;\;\;
u^+_1 - u^-_1  =  (u^+_L - u^-_L)/\epsilon,
\label{eq:jump1}\\
u^+_R + u^-_R  =  u^+_3,
\;\;\;
u^+_R - u^-_R  =  u^+_3/\epsilon,\;\;\mbox{($u^-_3=0$)},
\label{eq:jump2}
\ea
here labeled $u^\pm_{1,L,R,3}$. Superscripts $\pm$ indicate the
direction of propagation (rightward and leftward respectively), while $1,L,R,3$
specifies the position with respect to the T.R. where they are evaluated: left chromosphere, left
corona, right corona, right chromosphere respectively. We further use the
notation $ph,1$ and $ph,3$ for evaluations at the left and right footpoints
(see Fig.~\ref{fig:1}).\\
We chose solutions of the form 
$u \propto\exp[-i\omega t - t/\tau]$ 
in which
$\omega$ is the wave frequency and $\tau$ is a damping timescale. 
A wave of amplitude $u_{ph,1}^+=u_1^0$ is injected at the left footpoint so that
$u_1^+=u_1^0\exp[-i\omega t_{ph}]$.
Time is counted in coronal
crossing time $t=n t_c=n L_c/V_a^c$, $n$ indicating the number of 
reflections,
so
that wave propagation and damping in the corona (we exclude damping in the
chromosphere for reasons that will be clear after) are 
accounted by the factors
$\xi=\exp[-i\omega t_c]$ and 
$\beta=\exp[-t_c/\tau]$:
\ba
\left. u^+_R \right|_{n+1}=\xi\beta\left. u^+_L \right|_n,
\;\;\;\;
\left. u^-_L \right|_{n+1}=\xi\beta\left. u^-_R \right|_n. 
\label{eq:prop1}
\ea
Combining Eqs.~\ref{eq:jump1}-\ref{eq:prop1} during a cycle of reflection
and propagation in the corona yields a recurrence formula,
$a=(1-\epsilon)/(1+\epsilon)$,
\be
u^+_L|_n = (1+a)u^+_1|n + a^2\beta^2\xi^2 u^+_L|_{n-2}
\label{eq:rec}
\ee
(and a similar expression for $u^-_L$) that can be 
used
to evaluate for each variable $u^\pm_i$:
a) its \emph{asymptotic} value, setting $n=n-2$; 
b) its \emph{temporal} evolution, 
expliciting
$u|_{n-2}=f(u^+_1|_{n-4},~u|_{n-4})$ and so on.\\
Here
we consider only two special cases: a) Finite frequency fluctuations ($\xi\ne1$)
with no damping; b) Zero frequency fluctuation with damping ($\beta\ne1$).
\section {No damping, $\omega\ne0$}
\begin{figure}[t]
\centering
\includegraphics[width=1\columnwidth]{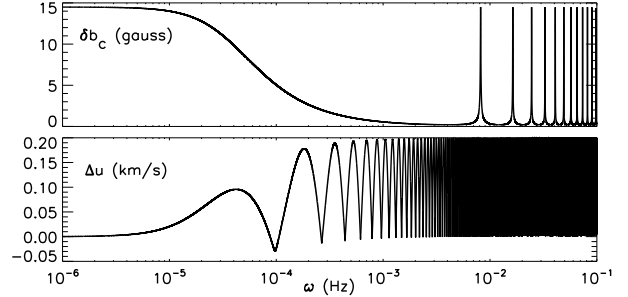}
\caption{Asymptotic shear between footpoints $\Delta u$ (bottom) and coronal
magnetic field (top) as a function of frequency.}
\label{fig:2}
\end{figure}

With $\beta=1$, the asymptotic solutions for the reflected and transmitted
wave can be written as
\ba
u^-_1 = a\frac{1-\xi^2}{(1-a^2\xi^2)}u^+_1,\;\;\;
u^+_3 = \frac{(1-a^2)\xi}{(1-a^2\xi^2)}u^+_1,\label{eq:asym1}
\ea
while in the corona one finds
\ba
u^+_L = \frac{(1+a)\xi}{(1-a^2\xi^2)}u^+_1,\;\;\;
u^+_R = \xi u^+_L\label{eq:asym2}\\
u^-_L = -a\frac{(1+a)\xi^2}{(1-a^2\xi^2)}u^+_1,\;\;\;
u^-_R = -a u^+_R\label{eq:asym3}
\ea
For injected waves of vanishing frequency and very low coronal density (small
$\epsilon$) one finds immediately from Eqs.~\ref{eq:asym1}-\ref{eq:asym3} that
$u^+_L\approx1/(1-a)u^+_1\approx1/\epsilon u^+_1\gg u^+_1$
and
$u^+_3=(1-a)u^+_R\approx2\epsilon u^+_R$.
Hence full transmission ($u^+_1\approx u^+_3$) implies very large coronal
amplitudes $u^\pm_c \approx1/\epsilon$ (larger than the WKB prediction
$u_{WKB}\approx1/\sqrt{\epsilon}$) and also $u^+_c\approx-u^-_c$. The latter
finally implies 
$\delta b/\sqrt{4\pi\rho}\gg\delta u$, that is magnetic energy much larger than
kinetic energy in the corona. The asymptotic solution is actually a uniform
$\delta u$ and $\delta b$ along the loop, while line-tying would have given
a linear profile for $\delta u$ (from $u_1^0$ to 0) and $\delta b$ that goes to
$\infty$ (never ending shear)\footnote{the line-tying can be recovered 
taking the limit $\epsilon\rightarrow0$ before $n\rightarrow\infty$}.\\
The asymptotic velocity shear between the right and left footpoints
$\Delta u = |\delta u_{ph,1}| - |\delta u_{ph,3}|$ and the coronal
magnetic field $\delta b_c$ for finite frequency fluctuations are shown in
Fig.~\ref{fig:2} as obtained from Eqs.~\ref{eq:asym1}-\ref{eq:asym3} assuming
a 10~Mm loop,  $L_{ph}=2~\mathrm{Mm}$, $V_a^{ph}=0.7~\mathrm{km/s}$,
and $\epsilon=1/70\ll1$.
$\delta b_c$ ranges from the photospheric value, $\delta
b_{ph}=15~\mathrm{Gauss}$, at zero frequency to zero at
high frequencies. 
Because of the almost perfect reflection at the two T.R.s the
coronal part of the loop behaves as a resonant cavity
(compare the poster \citep{Malara_al_2009} in the same session and
\citep{Nigro_al_2008}), producing peaks at the fundamental frequency
$V_a^c/L_c$ and its harmonics.
On the other hand the asymptotic shear (Fig.~\ref{fig:2}, bottom panel)
has well defined peaks well below such frequency that are caused by the phase
shift between the injected and reflected waves at the left footpoint, i.e. a
dependence of the form $\Delta u =g(\omega t_{ph})$,
$t_{ph}=L_{ph}/V_a^{ph}$ being the chromospheric crossing time.
Figure \ref{fig:3} shows the temporal evolution of $\Delta u$ for different
frequencies (black lines) for the same 10~Mm loop. 
The relaxation is independent of $\epsilon$ 
(cf. the plots for $\omega=0~\mathrm{Hz}$ obtained with $\epsilon=1/70$ and
$\epsilon=0.4$), which allows to derive a compact formula for $\Delta
u(t)$ by expliciting the dependence 
$u_{n-2}=f(u_{n-4})=f(u_{n-6})...$ and taking the limit $\epsilon\rightarrow0$:
\ba
u^-_1 &=& u^+_1 a\left[1-(1-a^2)\xi^2{\cal H}(t)\right]\\
u^+_3 &=& u^+_1 (1-a^2)\xi{\cal H}(t)
\ea 
\ba
\mbox{with};\;\
{\cal H}(t)=\frac{1-\exp\left[-2t/\tau_{rel}-i\omega t\right]}
{1+ i\omega\tau_{rel}}
\ea
and $\tau_{rel}=L_c/V_a^{ph}$ being the relaxation timescale
analogous\footnote{In their model atmosphere, \citet{Grappin_al_2008} have
non-vanishing \alfven speed gradients at the photospheres,
corresponding to our T.R..} to
that found by
\citet{Grappin_al_2008}, $\tau_{rel}=L/V_a^{ph}$.
The peak at $t\approx2t_{ph}$ corresponds to the initial acceleration of the
left footpoint exerted by the first reflection at the T.R. 
(transient line-tying). The
following relaxation is due to the deceleration of the left footpoint and the
acceleration of the right footpoint. Oscillations in $\Delta u(t)$ for finite
frequencies are caused by the phase shift between the injected and reflected waves at the left photosphere (the $i\omega t$ term).
\section{Turbulent damping, $\omega=0$}
\begin{figure}[t]
\centering
\includegraphics[width=1\columnwidth]{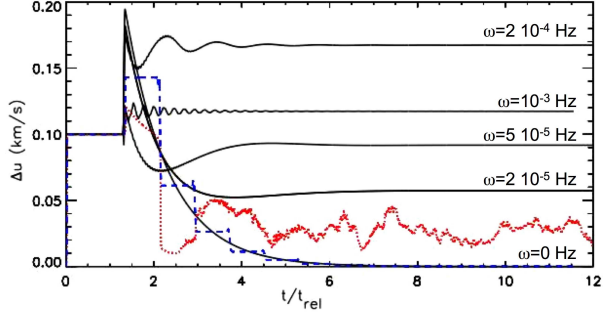}
\caption{Velocity difference between footpoints $\Delta u$ 
as a function of time for different frequencies, 
for $\omega=0$ and $\epsilon=0.4$ (blue-dashed), and for the weakly nonlinear case
(red-dotted).}
\label{fig:3}
\end{figure}
\begin{figure}[t]
\centering
\includegraphics[width=1\columnwidth]{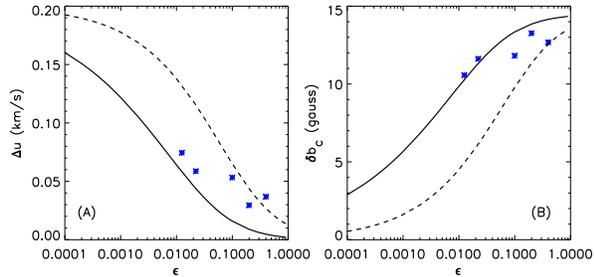}
\caption{Velocity shear $\Delta u$ (A) and 
asymptotic coronal magnetic field
$\delta b_c$ (B) as a function of $\epsilon$ from  
eq.~\ref{eq:timeev} and eq.~\ref{eq:ind} solved with different
nonlinear timescales: a Kolmogorov (dashed) and an IK
(solid) type. Symbols are results from the shell simulations.}
\label{fig:4}
\end{figure}

We are now interested in understanding whether the above relaxation (for zero
frequency) takes place in presence of a damping, that we chose to be caused by
incompressible turbulence triggered by counter-propagating waves. 
We perform simulations in which the nonlinear MHD
dynamics in planes perpendicular to the mean magnetic field is replaced by
two-dimensional shell models for $\delta u$ and $\delta b$. The
shell models of different planes are coupled by \alfven waves propagating
along the mean magnetic field \citep{Nigro_al_2004,Buchlin_Velli_2007}.
In each layer the damping factor is given (dimensionally) by 
$\beta=\exp[-t_{ph,~c}/\tau_{nl}^{ph,~c}]$, with
$\tau_{nl}^{ph,~c}=l_\bot/u_{ph,~c}^\pm$ and $l_\bot$ is the outer scale of turbulence taken to coincide with the loop width. 
We expect that
the amplitude in the corona saturates to value smaller than $u^\pm_c\propto1/\epsilon$ when subject to a
damping, say the WKB value $u_c\propto1/\sqrt{\epsilon}$. One gets 
$t_c/\tau_{nl}^c=(u_1^0\sqrt{\epsilon}/V_a^{ph})(L_c/l_\bot)$ that justifies
neglecting the damping in the chromosphere.
Based on this dimensional analysis, we set 
$u_1^0=0.2~\mathrm{km/s},~L_c=6~\mathrm{Mm}$
$\epsilon=0.4,~L_c/l_\bot=1,~u_1^0/V_a^{ph}=2/7$ in the simulation to obtain a
weak turbulence limit\footnote{Keeping fixed the ratio
$u_1^0/V_a^{ph}$, a different choice of the nonlinear timescale,
as done below, would give a slightly higher $\epsilon$.}. 
The velocity shear between footpoints relaxes 
in a few characteristic timescales, although a small shear survives, oscillating
with time (red-dotted curve in Fig.~\ref{fig:3}). For smaller $\epsilon$,
relaxation occurs earlier and to larger $\Delta u$ (cf. symbols in
Fig.~\ref{fig:4}, A), with oscillation on similar timescale
(not shown). The nature of the oscillations is not completely clear,
although it is related to the sharp reflection at the T.R. and
is expected to be reduced when the
T.R. is properly accounted for.\\
The asymptotic shear can be understood in our simple analytical picture
adopting a phenomenological expression for the turbulent damping.
With $\beta<1$ and $\xi=1$ in the recurrence formula, and in the limit of
small $\epsilon$ one obtains for the wave amplitude in the corona:
\ba
u_L^+=\frac{1}{2\epsilon}u_1^0\frac{1-\exp[-2t/\tau_{rel}-t/\tau_{nl}]}
        {1+\tau_{rel}/2\tau_{nl}}.
\ea 
At the beginning, when few reflections have occurred, $\tau_{nl}\gg\tau_{rel}$ and the
system evolves as in the linear case, with the wave amplitude in the corona 
growing toward the asymptotic non-WKB value $u^+_L=u_1^0/\epsilon$. 
When $\tau_{nl}\lesssim\tau_{rel}$ the
relaxation process stops and the level of fluctuation saturates to the value:
\ba
u_L^+=\frac{1}{2\epsilon}u_1^0\frac{1}
        {1+\tau_{rel}/2\tau_{nl}}
\label{eq:timeev}
\ea
Assuming that $u_L^+\approx u_L^-$ (or equivalently $\delta u \ll \delta
b_c/\sqrt{4\pi\rho_c}$), and choosing a 
Kolmogorov type phenomenology for the nonlinear timescale
\citep{Kolmogorov_1941},
$\tau_{nl}={l_\bot}/{u_L^+}$,
yields the asymptotic level of the coronal magnetic field
\ba
b_\infty=\frac{\delta b_c^\infty}{\sqrt{4\pi\rho_c}}
\approx \left.u_L^+\right|_\infty=u_1^0\frac{\tau_{nl}^{ph}}{\tau_{rel}}
\left[\sqrt{1+\frac{\tau_{rel}}{\tau_{nl}^{ph}}\frac{1}{\epsilon}}-1\right]
\label{eq:bKolm}
\ea
Given the value of $u^+_c\approx u^-_c$, the velocity shear at
footpoints can be obtained from the asymptotic expression or
the time evolution formula for $\Delta u$. Here we prefer to exploit the
stationary state of the system in the induction equation, which in a
dimensional form leads to
(other choices gives similar results):
\ba
\dt{b_\infty}\approx V_a^c\frac{\Delta u}{L_c}-\frac{b_\infty}{\tau_{nl}}=0,
\label{eq:ind}
\ea
from which one finally gets,
\ba
\Delta u = \frac{L_c}{l_\bot}\frac{b^2_\infty(\epsilon)}{V_a^{ph}}\epsilon
\label{eq:delta_kolm}
\ea
With a different choice of the nonlinear timescale, as in the IK phenomenology
\citep{Iroshnikov_1964,Kraichnan_1965},
$\tau_{nl}=l_\bot V_a^c /(u_L^+)^2$,
one obtains a cubic expression for $u^+_c$, which has to be solved numerically,
but the induction equation can still be used to get the velocity shear.
Note that for small $\epsilon$, Eq.~\ref{eq:bKolm} gives a dependence
consistent with the WKB estimate $u^\pm_c\propto\epsilon^{-1/2}$, 
while the IK phenomenology gives $u^\pm_c\propto\epsilon^{-2/3}$, that is a
coronal magnetic field intermediate between the WKB and linear solutions.
In Fig.~\ref{fig:4} symbols represent $\delta b_c$ and $\Delta
u$ computed at the coronal base obtained with the simulations (and averaging
over 4-5 oscillations), varying
$\epsilon$ while keeping fixed the other parameters (loop length and width,
\alfven speed at the photosphere, injected wave amplitude). The solid and
dashed lines are the analytical estimates obtained with a Kolmogorov and IK
phenomenology. A relatively good agreement is found between the simulations at high \alfven speed contrast (small
$\epsilon$) and the IK
phenomenology. For the latter a fiarly accurate approximation is given 
by (cf. Fig.~\ref{fig:4}, B for $\epsilon\lesssim10^{-3}$):
\ba
\delta b_c^{\infty}\approx \delta b^0_{ph}
\left[\frac{l_\bot}{L_c}
\left(\frac{V_a^{ph}}{u^0_1}\right)^2\epsilon\right]^{1/3},
\ea
($\delta b^0_{ph}=1/2 u_1^0\sqrt{4\pi\rho_{ph}}\approx15~\mathrm{Gauss}$).
For small reflection ($\epsilon\gtrsim0.02$), $u^+_c\nsim u^-_c$,
$\delta b_c/\sqrt{4\pi\rho_c}\nsim u^\pm_c$, so that
both phenomenologies fail to reproduce the
simulation results. 
\section{Discussion}
Adopting a simplified model based on \alfven wave propagation and reflection 
we recover the essential features found in \citet{Grappin_al_2008}. A loop 
driven at one footpoint by a ``zero-frequency'' photospheric velocity
relaxes to a 
uniform transverse velocity $\Delta u =0$ and magnetic field $\delta
b_c=\delta b_{ph}$ on a characteristic timescale 
$\tau_{rel}=L_c/V_a^{ph}$. This situation continues to hold basically for
$\omega\ll\omega_*\approx1/\tau_{rel}$. For $\omega\approx\omega_*$, 
the asymptotic shear is finite (partial relaxation) and 
$\delta b_c<\delta b_{ph}$. 
For $\omega\gg\omega_*$, our model correctly reproduces the behavior of the
coronal magnetic field that tends to 0 and shows resonances at well defined frequencies. 
On the countrary oscillations in the velocity shear, $0<\Delta u = u_1^0$, are 
exaggerated in our simplified model but when the T.R. is represented as a continuous
layer we recover the correct behavior $\Delta
u\rightarrow0$. The system is neither line-tied ($\delta u=0$) nor
relaxed ($\delta b_c=0$).
When weak (turbulent) damping is taken into account, both numerical simulations 
and analytical calculations show that the relaxation process still occurs. In the strong turbulence limit, the loop relaxation is stopped as 
soon as the coronal nonlinear time becomes smaller than the coronal 
propagation time ($\tau_{nl}^c< t_c$). 
The asymptotic magnetic field in the corona (in
velocity units) has a scaling that varies between $\epsilon^{-1/2}$ and
$\epsilon^{-2/3}$, depending on the phenomenology (Kolmogorov and IK
respectively, the latter showing a better agreement with the simulations), which is intermediate between the linear ($1/\epsilon$) and WKB
($1/\sqrt{\epsilon}$) scalings. 
The relaxation still occurs when the ratio $\tau_{nl}^c/\tau_{rel}>1$. According to the
definition of nonlinear timescale in the Kolmogorov or IK phenomenology, this can be
rewritten as 
$\tau_{nl}^c/\tau_{rel}=(\epsilon\tau_{nl}^{ph}/\tau_{rel})^{1/2}$
and 
$\tau_{nl}^c/\tau_{rel}=(\epsilon\tau_{nl}^{ph}/\tau_{rel})^{1/3}$
respectively. 
For a given loop width and initial footpoint shear, and assuming that longer
loop has higher density contrast (smaller $\epsilon$)
the asymptotic state is controlled by loop length.

\emph{Acknowledgments} This research was carried out in
part at the Jet Propulsion Laboratory, California Institute of Technology,
under a contract with the National Aeronautics and Space Administration. It was
also supported by the Italian Space Agency contract Solar System Exploration
and by the Belgian Federal Science Policy Office through the ESA-PRODEX program.
  \bibliographystyle{aipproc}

\end{document}